\documentclass[%
 aip,
 amsmath,amssymb,
 reprint,%
]{revtex4-1}

\usepackage{graphicx}
\usepackage{dcolumn}
\usepackage{bm}

\usepackage[utf8]{inputenc}
\usepackage[T1]{fontenc}
\usepackage{mathptmx}
\usepackage{etoolbox}
\usepackage{xcolor}
\usepackage{makecell}
\usepackage{hyperref}
\newcolumntype{K}[1]{>{\centering\arraybackslash}m{#1}}

\makeatletter
\def\@email#1#2{%
 \endgroup
 \patchcmd{\titleblock@produce}
  {\frontmatter@RRAPformat}
  {\frontmatter@RRAPformat{\produce@RRAP{*#1\href{mailto:#2}{#2}}}\frontmatter@RRAPformat}
  {}{}
}%
\makeatother
\begin{document}

\preprint{AIP/123-QED}

\title[A Transition Edge Sensor Operated in Coincidence with a High Sensitivity Athermal Phonon Sensor for Photon Coupled Rare Event Searches]{A Transition Edge Sensor Operated in Coincidence with a High Sensitivity Athermal Phonon Sensor for Photon Coupled Rare Event Searches}

\author{R.K. Romani} \thanks{Corresponding author: \href{mailto:rkromani@berkeley.edu}{rkromani@berkeley.edu}}\affiliation{University of California Berkeley, Department of Physics, Berkeley, CA 94720, USA}

\author{Y.-Y. Chang} \affiliation{University of California Berkeley, Department of Physics, Berkeley, CA 94720, USA}

\author{R. Mahapatra} \affiliation{Texas A\&M University, Department of Physics and Astronomy, College Station, TX 77843-4242, USA}
\author{M. Platt} \affiliation{Texas A\&M University, Department of Physics and Astronomy, College Station, TX 77843-4242, USA}
\author{M. Reed} \affiliation{University of California Berkeley, Department of Physics, Berkeley, CA 94720, USA}
\author{I. Rydstrom} \affiliation{University of California Berkeley, Department of Physics, Berkeley, CA 94720, USA}
\author{B. Sadoulet}\affiliation{University of California Berkeley, Department of Physics, Berkeley, CA 94720, USA}
\author{B. Serfass} \affiliation{University of California Berkeley, Department of Physics, Berkeley, CA 94720, USA}
\author{M. Pyle} \affiliation{University of California Berkeley, Department of Physics, Berkeley, CA 94720, USA}

 \email{rkromani@berkeley.edu.}

\date{12/05/24}

\begin{abstract}
Experimental searches for axions or dark photons that couple to the standard model photon require photosensors with low noise, broadband sensitivity, and near zero backgrounds. Here, we introduce an experimental architecture, in which a small photon sensor, in our case a Transition Edge Sensor (TES) with a photon energy resolution $\sigma_\gamma = 368.4 \pm 0.4$ meV, is colocated on the same substrate as a large high sensitivity athermal phonon sensor (APS) with a phonon energy resolution $\sigma_\mathrm{phonon} = 701 \pm 2$ meV. We show that single 3.061 eV photons absorbed in the photon-sensing TES deposit $\sim$35\% of their energy in the electronic system of the TES, while $\sim$26\% of the photon energy leaks out of the photon-sensing TES during the downconversion process and becomes absorbed by the APS. Backgrounds, which we associate with the broadly observed ``low energy excess'' (LEE), are observed to be largely coupled to either the TES (``singles'' LEE), or phonon system, (``shared'' LEE).  At high energies, these backgrounds can be efficiently discriminated from TES photon absorption events, while at low energies, their misidentification as photon events is well modeled. With significant sensitivity improvements to both the TES and APS, this coincidence technique could be used to suppress backgrounds in bosonic dark matter searches down to energies near the superconducting bandgap of the sensor. 
\end{abstract}

\maketitle

Multiple tests of Beyond the Standard Model (BSM) physics search for the coupling of proposed BSM particles to an optical or infrared standard model (SM) photon. In ``dish antenna'' type searches for axion dark matter \cite{DishAntennaConcept} (e.g. BRASS \cite{BRASS} and BREAD \cite{BREAD}), high magnetic fields coupled to reflectors attempt to convert dark matter halo axions or axion-like-particles to SM photons via the inverse-Primakoff effect. In contrast, ``light shining through walls'' type experiments (e.g. ALPS \cite{ALPS}) convert SM photons to axions or axion-like-particles through the Primakoff process before converting them back into SM photons through the inverse-Primakoff effect. Finally, other experimental architectures search for photon conversion in dielectric stacks (e.g. LAMPPOST\cite{LAMPPOST} and MADMAX\cite{MADMAX}). 

In all three experimental architectures, tests of BSM physics rely on the ability to search for rare photons using high performance photon sensors. Such sensors would ideally have a large dynamic range (to test the maximum possible particle mass/energy range), low noise (to search for the smallest energy photons), and near zero background rate that is indistinguishable from the BSM signal. We summarize the state of the art for several sensor technologies in Table \ref{tab:comparison}. 

Transition Edge Sensors (TESs) \cite{irwinTransitionEdgeSensors2005} are a mature, low noise, relatively large dynamic range sensor technology that has been proposed for use in such photon coupled rare event searches \cite{BREAD, ALPS-II} and demonstrated as astronomical sensors \cite{CabreraOpticalTES, Hattori_2022}. TESs lead in applications where good energy resolution and low background rates are desired, making them well suited to photon coupled rare event searches. 

However, recent high-performance TES based dark matter calorimeters have measured a background rate of events that seem to originate in the sensor \cite{CRESSTDoubleTES, TwoChannelPaper}. These events (frequently labeled ``singles'' because of their coupling to a single sensor channel) are thought to be part of a broader class of backgrounds often called the ``Low Energy Excess'' (LEE) \cite{adariEXCESSWorkshopDescriptions2022}, which have a characteristic rising rate of events below $\sim$ 200 eV and have been observed to fall in rate with time\cite{CRESSTTimeVariationLEE, anthony-petersenStressInducedSource2022, EDELWEISSTimeDependence}, and which severely limit the science reach of light dark matter direct detection calorimeters and coherent neutrino scattering (CE$\nu$NS) experiments. While the origin (or origins) of LEE events is still unknown, they have been broadly hypothesized to originate from the relaxation of mechanical stress \cite{anthony-petersenStressInducedSource2022, AlRelaxation}. In photon coupled rare event searches, LEE ``singles'' would mimic events expected from e.g. axions converting to SM photons, constituting a limiting background.

In this Letter, we propose and demonstrate a detector architecture designed to measure both the localized energy deposited in the TES electronic system and the phonon energy dropped into the substrate by a photon absorbed by the TES. Requiring coincidence of these two signals discriminates real TES photon absorption from both substrate phonon backgrounds (e.g. ``shared LEE'', radiogenic gamma rays absorbed in the substrate) and backgrounds that entirely deposit their energy in the sensor electronic system (``singles LEE''). Note that our architecture is not inherent to the use of TESs, different sensors (KIDs\cite{PhononKID, MazinDevice}, Qubit-derived sensors\cite{QCD, SQUAT}, etc.) could be substituted for our photon or phonon sensors.

Photon interactions with TESs are commonly modeled by assuming that the photon initially transfers all its energy to a single electron, which quickly downconverts to a large number of excited electrons (i.e. quasiparticles) and phonons. \cite{KozorezovQPs} This electron-phonon cascade process ultimately results in an elevated thermal energy in the TES electronic system that is responsible for the TES response. During this downconversion process, energy is lost when phonons radiated from excited electrons have low enough energy to escape the TES into the substrate. Previous works\cite{Cabrera2005, KozorezovQPs, Kozorezov2007, Kozorezov2008, goldie1994statistical, Rando1992} have indicated that in TES systems of this type the photon energy is roughly equally partitioned between the TES electronic system and the phonons lost to the substrate, broadly consistent with previous measurements of the fraction of optical photon energy absorbed locally in TESs \cite{CabreraOpticalTES}. However, the fraction of radiated phonons that can break aluminum Cooper pairs and thus be sensed by superconducting athermal phonon sensors has never been measured. 

\begin{figure}
\includegraphics[width=1\columnwidth]{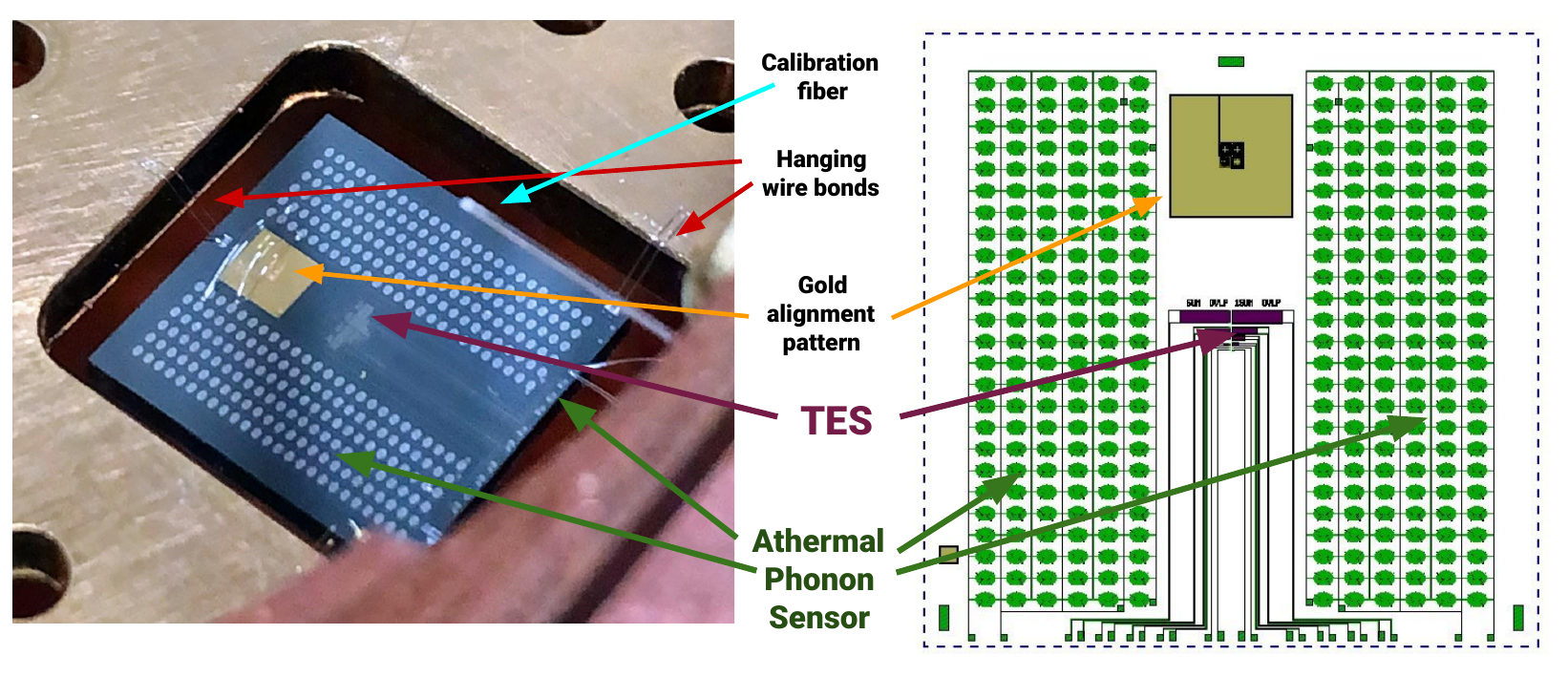}
\caption{\label{fig:detector_pic} (Left) Photograph of our device mounted in the cryostat. Note that the device is suspended from wire bonds (see text), and illuminated from above by a fiber with a diffusing tip. (Right) The design of the device. Note that only the top right (i.e. 200 $\mu$m by 800 $\mu$m) TES is read out for this study.}
\end{figure}

Here, we introduce a 200 $\mu$m $\times$ 800 $\mu$m $\times$ $\sim$40 nm  65.1 $\pm$ 0.2 mK effective Tc W TES with a $\sigma_\mathrm{TES}$ = 128.5 $\pm$ 0.1 meV energy resolution for Dirac-delta energy depositions within the TES, patterned on a 1 cm x$\times$ 1 cm $\times$ 1 mm thick silicon substrate instrumented with an Athermal Phonon Sensor (see Fig. \ref{fig:detector_pic}). An Athermal Phonon Sensor (APS) is constructed by coupling an array of W TESs to aluminum athermal phonon collection fins in the common QET architecture \cite{irwinQuasiparticleTrapAssisted1995}.  Approximately $64.3 \%$ of the phonon-absorbing metal on the device surface is composed of the active APS, while the TES used to absorb optical photons makes up only $\sim 1.07 \%$ of the metal on the device surface. Because of this difference in coverage fraction, photon events absorbed in the substrate will predominantly couple to the APS; very little response is expected in the TES. Our device is suspended by wire bonds as in Ref. \cite{anthony-petersenStressInducedSource2022} in order to suppress ``Low Energy Excess'' (LEE) backgrounds associated with gluing the substrate to a copper mount. By comparing the response in these two channels (the photon sensing TES and the phonon sensing APS), we aim to discriminate photons hitting the TES from background events.

\begin{figure}
\includegraphics[width=1\columnwidth]{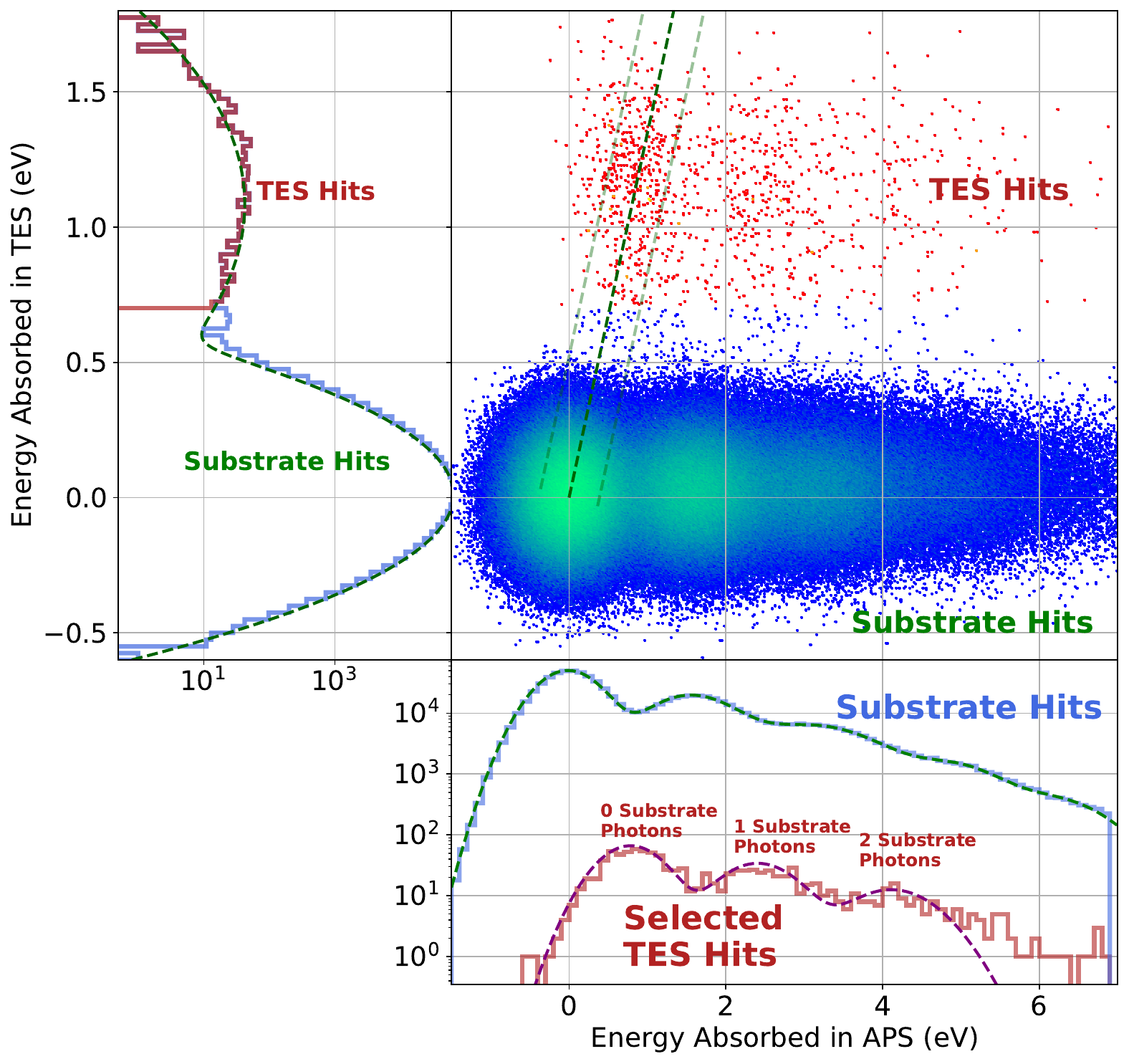}
\caption{\label{fig:calibration} (Main) Heatmap plot of the energies absorbed in the TES and APS from 3.061 eV photon calibration events, each point corresponds to the device response from one photon calibration pulse. Most photons hitting the device are absorbed in the substrate phonon system, leading to large response in the APS with very little response in the TES (blue/green points). Rarely, photons directly hit the TES, creating a large response (red points). Dashed dark and light green lines (derived from equations \ref{eqn:e_gamma} and \ref{eqn:chi2}) show the expected mean (dark green, dashed) response and the band in which most events are expected to fall (light green, dashed, $\chi^2_\gamma = 1$) for events where a photon is absorbed in the TES. Most events corresponding to a single photon absorbed in the TES fall within the expected band, consistent with expectations. (Left) Histogram of energies absorbed in the TES. (Bottom) Histogram of energies in the APS for all events (blue) and events which are tagged as TES hits (red). Dashed lines show Gaussian fits. }
\end{figure}

To calibrate the device response, we illuminate the entire device surface with pulses of 405 nm (3.061 eV) photons with an average of $\lambda = $ 0.165 $\pm$ 0.012 photons per bunch (see Fig. \ref{fig:calibration}). We trigger in coincidence with the laser pulse, and measure the response in our TES and APS channels using an optimal/matched filtering approach (see Fig. \ref{fig:calibration}). As we trigger at the time of the laser pulse regardless of the signal in the TES or APS, the energy estimator at this trigger time may fluctuate randomly upward or downward from the ``true'' event energy, given this energy estimator convolves the true response of the sensor with the sensor's approximately Gaussian noise distribution function. For laser pulses in which no photons happen to hit the device, we expect and observe a Gaussian distribution of fit pulse heights centered at zero.

Most calibration photons are absorbed in the substrate phonon system, and are used to calibrate the response of the APS. In the APS, we measure a phonon energy resolution of $\sigma_\mathrm{phonon} = $ 701 $\pm$ 2 meV and a phonon collection efficiency of $\epsilon_\mathrm{phonon} = $ 0.499 $\pm$ 0.001 for substrate absorbed photons, determined by dividing the average energy absorbed in the APS for one photon events by the photon energy.

In rare cases, a calibration photon directly hits the TES, producing a large response. These ``TES direct hit'' events can be used to study the response of the TES to optical photons and to gain insight into the downconversion process occurring in the TES. In events with a TES direct hit, additional photons present in the calibration pulse may also interact with the substrate, creating a superposition of ``direct hit'' and ``substrate'' responses.

\begin{figure}
\includegraphics[width=0.95\columnwidth]{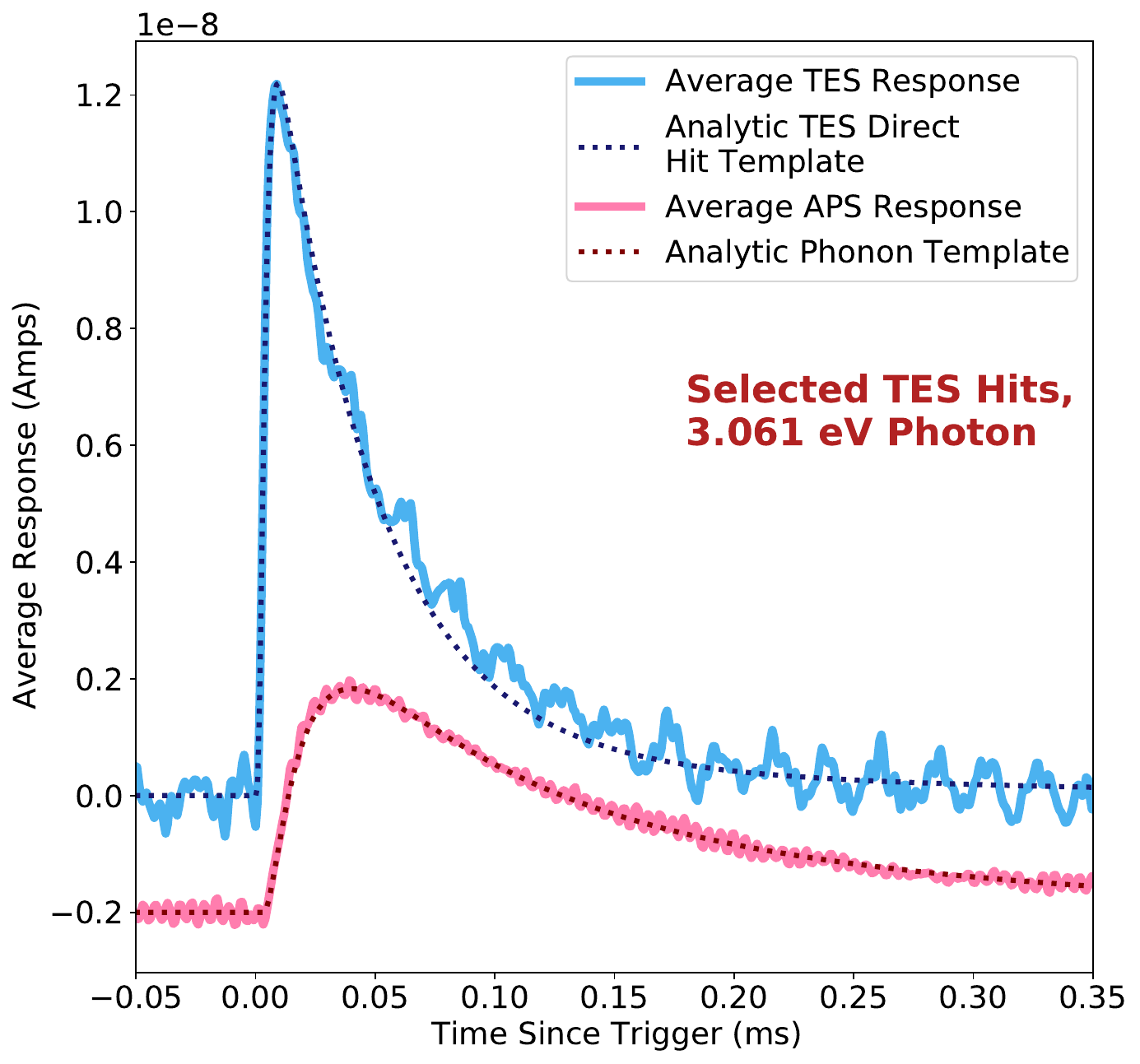}
\caption{\label{fig:response} The average time domain response in the TES (blue) and APS (pink) channels for events where a single photon is absorbed in the TES and no photon is absorbed in the substrate phonon system (events labeled ``0 Substrate Photons'' in Fig. \ref{fig:calibration}). Note the nonzero response in the APS. For clarity, averaged responses are low pass filtered ($f_{cut} = 100$ kHz) to remove high frequency electronics noise, and responses are baseline-subtracted and offset. For comparison, dotted lines show modeled responses for Dirac-delta energy impulses in the TES (dark blue, dotted) and phonon system events in the APS (dark red, dotted). }
\end{figure}

After selecting events with a large response in the TES, we plot a histogram of coincident energies seen in the APS. We fit this histogram with multiple Gaussians, corresponding to events with zero, one, two, etc. photons being absorbed in the substrate phonon system (in addition to the photon being absorbed in the TES). Notably, the zeroth peak is \textit{not} consistent with zero response in the phonon sensor. Selecting events from this zeroth peak, and comparing their average pulse shape to templates (see Fig. \ref{fig:response}), we see that the response in the TES channel is consistent with the modeled device response from a Dirac-delta-like energy deposition in the TES, consistent with a downconversion process which is fast compared to the $ O(10\mu s)$ TES response time. This response is modeled by convolving the TES responsivity $\frac{\partial P}{\partial I}(f)$ (modeled\cite{irwinTransitionEdgeSensors2005, ThreePoledPdI, watkinsThesis} from the measured TES complex impedance $dV/dI(f)$) with a Dirac delta function. By contrast, the response in the APS is consistent with the slow absorption of athermal phonons into metal films on the device surface (i.e. identical to calibrated substrate photon events). We interpret these phonons as originating from the leakage of athermal phonons from the TES during the electronic downconversion process following the absorption of a photon.

By fitting events in this peak (one photon absorbed in the TES, zero photons absorbed in the substrate), we measure the average partitioning of energy in the downconversion process. The energy absorbed in our TES and APS for each event can be calculated by multiplying the fit optimum filter amplitude $I_{OF}$ in units of current through the TES by a conversion factor $\frac{\partial E}{\partial I}$. This conversion factor was calculated using the normalized current domain event template $T_i(t)$ and the responsivity of the TES $\frac{\partial P}{\partial I}(f)$ by using Fourier and inverse Fourier transforms $\mathcal{F}$ and $\mathcal{F}^{-1}$ to calculate the power domain template $T_p(t)$: 
\begin{eqnarray}
    \frac{\partial E}{\partial I} = \int T_p(t) dt =  \int {\mathcal{F}}^{-1} \bigg( \frac{\partial P}{\partial I}(f) \mathcal{F}(T_i(t)) \bigg) dt
\end{eqnarray}

Using this procedure, we measure the energy absorbed in both the TES and APS, and find that for a $E_\gamma=$3.061 eV calibration photon direct hit on the TES, on average $\epsilon_T = E_{\mathrm{TES}}/E_\gamma = $ 35.1 $\pm$ 0.6 $\%$ is absorbed in the TES, while $\epsilon_V  = E_{\mathrm{APS}}/E_\gamma=$  26.4 $\pm$ 0.8 $\%$ is absorbed in the APS. After accounting for photon collection efficiency, we infer a baseline resolution of $\sigma_\gamma =$ 368.4 $\pm$ 0.4 meV for photons incident on the TES. In contrast, selecting events consistent with photon absorption in the substrate only we find that 49.9 $\pm$ 0.1 $\%$ of the photon energy is collected in the athermal phonon sensor and $\lesssim$ 1 \% is absorbed in the TES.

If we assume the phonon collection efficiency factor measured for substrate absorbed photons also applies to the phonons escaping the TES after a direct photon hit, we infer that $\sim$ 52.8 $\%$ of the initial photon energy is put into the substrate phonon system as phonons with energy above the superconducting aluminum bandgap. However, the phonon spectrum created by the downconversion of phonons absorbed in the substrate phonon system may differ from the spectrum from events absorbed in the TES (as e.g. the absorption of phonons within the TES would be expected to bias the phonon spectrum toward lower energies); therefore this estimated efficiency is only qualitative.

\begin{figure}
\includegraphics[width=1.0\columnwidth]{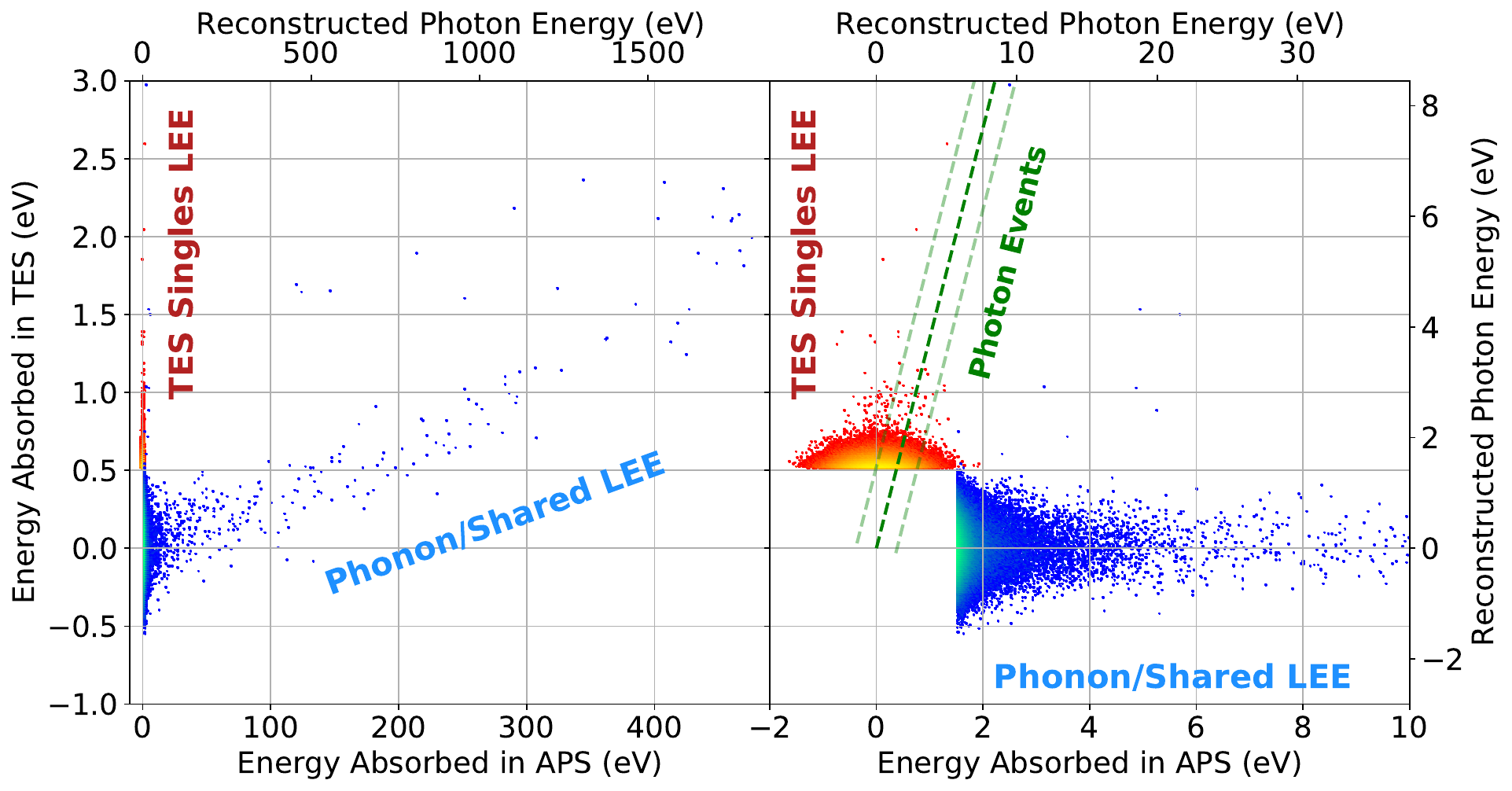}
\caption{\label{fig:backgrounds} The backgrounds observed in the device, triggered either on a signal in the TES (red) or in the APS (blue/green). Backgrounds triggered in the APS are consistent with phonon coupled Low Energy Excess (LEE). Backgrounds triggered in the TES, which we interpret as LEE ``singles'' seen in other devices, are consistent with a TES response from a impulse like energy deposition in the TES with little or no phonon response. This is in contrast to the expected response for photons, for which we expect an appreciable phonon response, following the dark green dashed line. Most TES photon absorption events will fall within the light green dashed lines (at $\chi^2_\gamma = 1$ of the mean expected response). Equations \ref{eqn:e_gamma} and \ref{eqn:chi2} give the relationship between the energy in the TES and APS channels for photon-like events with a given goodness of fit ($\chi^2_\gamma$) value. (Right) is a cropped view of (left).}
\end{figure}

Additionally, we collected 24h of background data in the device, triggering on the time domain optimal filtered\cite{MatchedFilters, sunilThesis, watkinsThesis} trace in both the TES and APS channels (see Fig. \ref{fig:backgrounds}). In this background dataset, we observed two main background types: phonon coupled backgrounds triggered in the APS, and ``singles'' triggered in the TES. These observations broadly agree with previous background measurements in similar detectors of LEE type backgrounds. \cite{TwoChannelPaper, CRESSTDoubleTES}

For events triggered in the APS, we observe a rising background below $\lesssim$ 500 eV which we attribute to phonon coupled LEE events, while at higher energies we see saturated events which we attribute to muon- and radioactive-decay-induced high energy backgrounds. From the coupling of these ``shared'' or phonon coupled LEE background events to the TES and APS, we can estimate that the TES collects approximately 1$\%$ of the phonon energy collected by the APS. We observed that in a given energy bin, the rate of low energy phonon backgrounds relaxes away with time. This time dependence is roughly consistent with either an exponential decay with an approximately weeklong time constant, or a $\sim 1/t$ time dependence, along the lines of the time dependence observed in Ref. \cite{anthony-petersenStressInducedSource2022}. If a $\sim 1/t$ time dependence is assumed, this may be consistent with the observed LEE time dependence in previous work by the CRESST and EDELWEISS collaborations \cite{CRESSTTimeVariationLEE, EDELWEISSTimeDependence}.

In the TES, we observe a spectrum that is consistent with random Gaussian noise below $\sim$0.75 eV. Above this energy, we observe a non-Gaussian outlier tail which has a pulse shape consistent with Dirac-delta impulses of energy deposited in the TES that we associate with LEE ``singles.''\cite{CRESSTDoubleTES, TwoChannelPaper} We did not observe a significant decrease in the rate of singles over time, but our ability to observe time dependence may have been limited by the large rate of noise events (discussed below) or the approximately weeklong datataking campaign.

The majority of these events are inconsistent with a photon-absorption-like energy partitioning between the TES and APS, leading us to discard photon backgrounds (see e.g. Ref. \cite{EssigBackgrounds}) as the primary cause of LEE singles seen in our devices. Additionally, the lack of a phonon component for the majority of events indicates that any process in the TES that results in a phonon or electron downconversion cascade from energies much greater than the superconducting bandgap is not a viable explanation for the majority of singles LEE events. A likely explanation for these events is bursts of sub aluminum bandgap photons that couple to the TES via  its antenna-like bias lines. In the future, we plan to aggressively filter the TES bias lines throughout the MHz-100GHz range such that such hypothetical external EMI bursts are significantly attenuated.

\begin{figure}
\includegraphics[width=1\columnwidth]{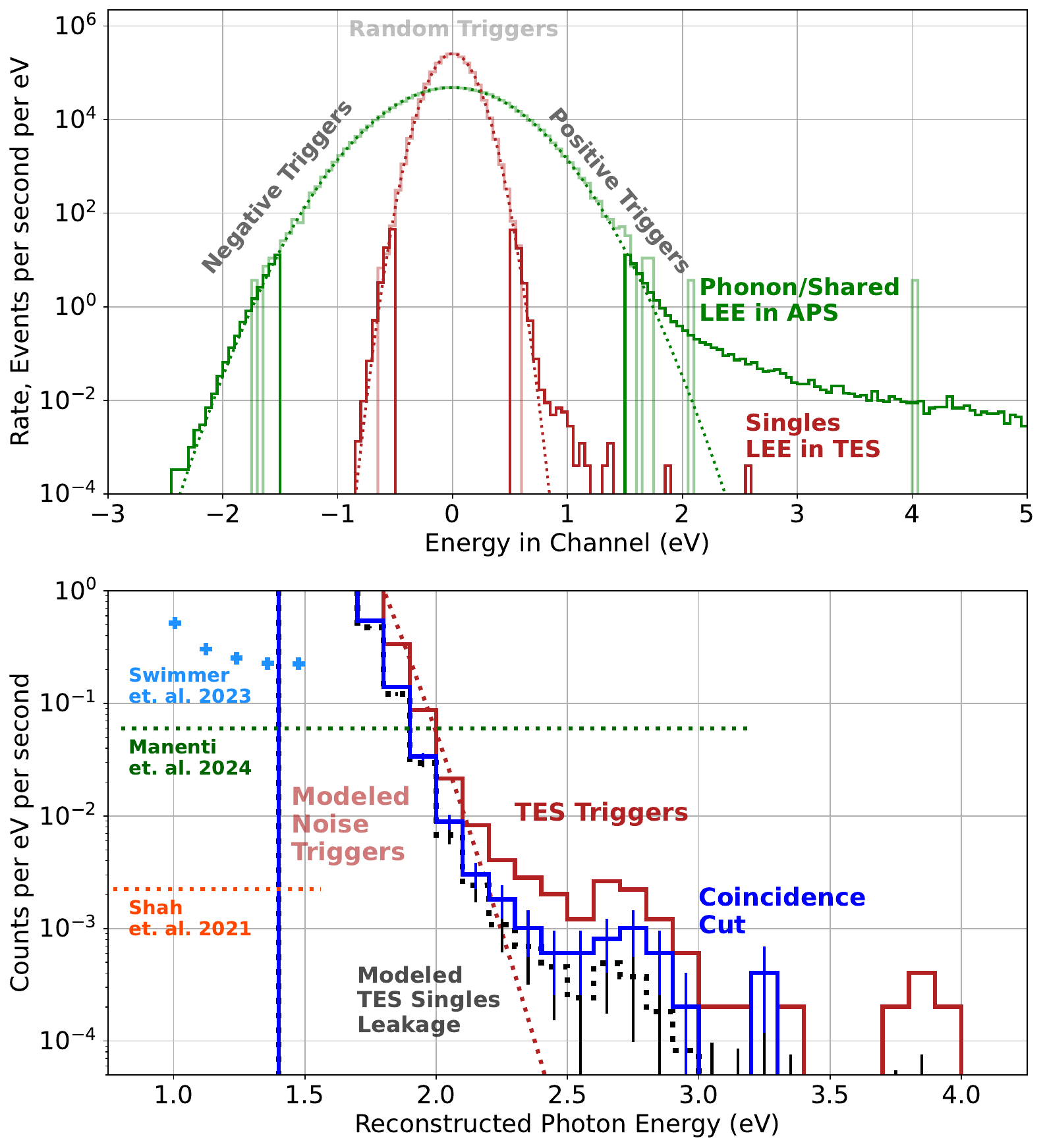}
\caption{\label{fig:spectra} (Top) Spectra of background events observed in either the APS (green) or TES (red). Solid spectra show triggers on positive or negative (noise sampling) amplitude events of significance $ \geq 4 \sigma$ in either the APS (green) or TES (red). Solid lighter color spectra show the reweighted event amplitudes of randomly triggered events which sample the noise. Dotted lines show Gaussian fits to the negative triggers, demonstrating that random fluctuations around zero describe the observed rate of events below $\sim$ 1.7 eV in the APS and $\sim$ 0.8 eV in the TES channel. (Bottom) Spectrum of photon-equivalent energies constructed through a weighted sum of energies in the TES and APS (see text) before (red) and after (dark blue) a coincidence cut requiring a response in both channels within $\chi^2_\gamma = 1$ of a photon-like response (see text). We model leakage of TES LEE singles into the acceptance band by assuming Gaussian smearing (black dotted histogram, see text), and find that this leakage largely explains the observed number of events passing our cuts. The red dotted line shows expected rate of noise triggers. Blue, green and orange dots show the background rates measured in Refs. \cite{MazinBackgrounds, TESBackgrounds, ALPSBackgrounds}, rescaled by sensor area. \cite{MazinDevice, ALPSDevice}}
\end{figure}

Despite the presence of two unmodeled backgrounds (shared and singles LEE \cite{CRESSTDoubleTES, TwoChannelPaper}), these backgrounds can be largely cut by requiring a coincident response in the TES and phonon channels consistent with a photon being absorbed by the TES (see Fig. \ref{fig:spectra}). Using inverse variance weighting, we reconstructed a photon energy
\begin{eqnarray}
\label{eqn:e_gamma}
    E_\gamma = \frac{1}{\sigma_{\gamma T}^{-2} + \sigma_{\gamma A}^{-2}} \bigg(\frac{E_{\gamma T}}{\sigma^2_{\gamma T}} + \frac{E_{\gamma A}}{\sigma^2_{\gamma A}}  \bigg) \\
    = \frac{1}{\sigma_{T}^{-2} \epsilon_T^2 + \sigma_{A}^{-2} \epsilon_A^2} \bigg(\frac{E_{T}/\epsilon_T}{\sigma^2_{\gamma T}/\epsilon_T^2} + \frac{E_{A}/\epsilon_A}{\sigma^2_{A}/\epsilon_A^2}  \bigg)
\end{eqnarray}
where $E_T$ ($\sigma_T$) and $E_A$ ($\sigma_A$) are the energies absorbed (resolutions) in the TES and APS respectively. $\epsilon_T$ and $\epsilon_A$ are the TES and APS collection efficiencies for photons absorbed in the TES (see above), determined by dividing the average energy absorbed in the TES and APS respectively by the energy of the photon incident on the TES. We also define a goodness-of-fit value for photon events
\begin{eqnarray}
\label{eqn:chi2}
    \chi^2_\gamma = \frac{\Big( E_{\gamma T} - E_{\gamma A} \Big)^2}{\sigma^2_{\gamma T} + \sigma^2_{\gamma A}}
\end{eqnarray}
These equations give the shape of the solid ($\chi^2_\gamma = 0$) and dashed ($\chi^2_\gamma = 1$) lines in Fig. \ref{fig:backgrounds}.  

Using this statistic, we can define a coincidence cut $(\chi^2_\gamma < 1)$ designed to preferentially cut events not consistent with a photon-like energy depositions. This is equivalent to accepting events within the green dashed lines in Fig. \ref{fig:backgrounds}. Through Gaussian smearing in the APS, a subset of triggers on LEE single events in the TES may appear to have APS responses consistent with photon absorption in the TES. Assuming that all TES singles have APS responses drawn from a Gaussian centered at zero, we model this leakage (dotted black line, Fig. \ref{fig:spectra}, bottom panel) by weighting each event seen in the TES by a factor derived from the probability it would randomly pass the coincidence cut. We show that in each bin, the observed photon-like events are consistent with Gaussian leakage of TES singles. Considering all bins, there appears to be a systematic bias toward observing more events than expected from LEE singles leakage, possibly suggesting two components of the singles background: a dominant background (gigahertz-scale EMI, stress relaxation) which drops very little energy into the phonon system, and a subdominant population of photon events that create the expected response. Such a population of photons might be caused by the scintillation of e.g. PCBs used to read out the device\cite{EssigBackgrounds}. Alternatively, a process which on average deposits a small phonon component into the substrate could also be responsible for observations.

In Fig. \ref{fig:spectra} we compare our device to benchmark TES\cite{TESBackgrounds, ALPSBackgrounds} and MKID\cite{MazinBackgrounds} photon sensors with reported background rates. We have  scaled the rates by the area of the sensor\cite{MazinDevice, ALPSDevice}, as many of the hypothetical sources of dark counts (stray environmental photons from insulators in the optical cavity, internal stress relaxation) would scale with the sensor area. In fact, we used a large 200 $\times$ 800$\mu$m$^2$ TES ($\sim$ 3 orders of magnitude larger than a standard optical TES) to enhance these volume scaled dark count rates. At high energies, our device compares favorably or comparably to Refs. \cite{TESBackgrounds, MazinBackgrounds, ALPSBackgrounds} (assuming the backgrounds in Refs. \cite{MazinBackgrounds, ALPSBackgrounds} remain flat to higher energies). At lower energies, the large size of our TES creates additional thermal fluctuation noise which unfortunately limits our threshold.  As discussed below, simply reducing the size and $T_c$ of our TES (as our group has already demonstrated\cite{finkTES}) would be expected to significantly reduce both threshold and background level.

In this Letter, we propose and demonstrate a optical photon detector architecture that reduces the impact of sensor-specific backgrounds by requiring that signals in a photon sensing TES be coincident with a pulse of athermal phonons created in the downconversion process in the TES. Our observation of this process supports long-standing models \cite{CabreraOpticalTES, Cabrera2005, KozorezovQPs, Kozorezov2007, Kozorezov2008, goldie1994statistical, Rando1992} describing electronic thermalization in superconducting and metal films. Our architecture is ideal for photon coupled rare event searches, which search for rare photons created by BSM processes.

\begin{table}
\begin{centering}
\caption{\label{tab:comparison} Comparison of our device to world leading TESs and other low temperature photon sensing technologies. World leading background rates and thresholds (5 $\sigma$ in the case of energy sensing sensors) are not necessarily achieved in the same device. For background rates, we normalize by the energy bin width and device area, as background rates from environmental photons or stress relaxation within the sensor would be expected to scale with the sensor area. Note that the high thresholds of our sensor are not inherent to our architecture; world leading sensors could be operated in similar devices.}
\begin{tabular}{|K{0.14\linewidth}|K{0.20\linewidth}|K{0.31\linewidth}|K{0.28\linewidth}|}
\hline
Sensor & Photon Energy Threshold (5$\sigma_\gamma$) [meV] &  Background Rate [Hz/(eV mm$^2$)] (Energy Range [eV]) & Notes \\ 
\hline
TES (this work) & 1840 & $3.7\times10^{-3}$ (2.3-2.9) & With discrimination \\
\hline
TES & 284\footnotemark[2] &  & Best demonstrated in our lab \\
\hline
\hline

TES (leading) & 142 \cite{Hattori_2022} & $1.4\times10^{-2}$ (0.8-1.5) \cite{ALPSBackgrounds} & Good threshold and backgrounds \\
\hline
KID (leading) & 816 \cite{Hattori_2022} & 1.41 (1.36-1.48) \cite{MazinBackgrounds} & Highly multiplexable \\
\hline
SNSPD (leading) & 43 \cite{SNSPDThreshold} & 2.7 $\times 10^{-5}$ (0.73-5.0)\footnotemark[1] \cite{LAMPPOST} & Not energy resolving \\
\hline
QCD (leading) & 6.2 \cite{QCD} & 6.2 $\times 10^{3}$ (0.006-5.0)\footnotemark[1] \cite{QCD} & Best demonstrated threshold \\
\hline

\end{tabular}
\end{centering}
\footnotetext[1]{No upper limit to energy sensitivity was given in Refs. \cite{LAMPPOST, QCD}. To normalize the background rate, we arbitrarily choose 5 eV.}
\footnotetext[2]{Estimated assuming a $\epsilon_T$ = 35.1 $\%$ as measured in this work. This device was an improved version of the device we published in Ref. \cite{finkTES}.}
\end{table}

Future versions of this device architecture should be designed to optimize two performance metrics: the photon energy resolution and the misclassification of TES singles as photon events. Our group has already demonstrated\cite{finkTES} TESs with significantly improved sensor resolution ($\sigma_{\mathrm{TES}}$ = 40 vs 128.5 meV) solely through decreasing our TES volume (100 $\times$ 400 $\mu$m $\times$ 40 nm vs 200 $\times$ 800 $\mu$m $\times$ 40 nm) and $T_c$ (40 vs 65 mK), and no fundamental limitations prevent the operation of world leading TESs\cite{Hattori_2022} in devices of this type. Indeed, simply scaling our TES to (25 $\mu$m)$^2$ and 40 mK would be expected to reduce $\sigma_\gamma$ to a world leading $\sim$ 11 meV, essentially eliminating the random noise that limits the performance of this specific device in the mid and near IR range. More efficient vetoing of TES singles requires improving the APS energy resolution in order to tag the relatively small phonon signal originating from photons being absorbed in the photon sensing TES. Our group has already demonstrated APSs with two times better phonon resolution\cite{TwoChannelPaper} than the APS in this paper by decreasing the quantity of instrumented and non-instrumented metal coverage on the substrate surface. Furthermore, the noise of the APS is completely dominated by LEE shot noise (sub threshold events which blur together into noise)\cite{TwoChannelPaper}, putting large sensitivity improvements within reach if LEE reductions can be achieved. However, we grant that the potentially superior performance of other superconducting photon and phonon sensor technologies (e.g. KIDs\cite{PhononKID}, qubit-derived sensors \cite{QCD, SQUAT}) may be needed to realize this coincidence tagging architecture in the far-IR where the threshold requirements on the photon and phonon sensors are more stringent. 

Finally, our results shed light on the nature of LEE singles and shared events. The majority of TES only, singles events must not primarily originate from a small number of high energy electrons or photons in the TES, as we do not see a coincident phonon pulse that leaks from the TES during the downconversion process; gigahertz scale photon bursts transmitted to the TES through the TES bias lines are a reasonable hypothesis that matches this observation. To evaluate the gigahertz EMI hypothesis we plan to filter our bias lines to attempt to remove this excess event source. We likewise plan to test the origin of the photon-like backgrounds in the TES (which pass our $\chi_\gamma^2$ cut) by minimizing the volume of scintillating materials in the device cavity that might stochastically emit photons.

Less can be gleaned on the source of the ``shared'' LEE phonon bursts. If the sole source of the phonon LEE events was stress relaxation in W films, we would expect roughly 14$\%$ of phonon LEE events to be associated with a large energy deposition in the monitored TES (given the TES is roughly $14 \%$ of the W on the device surface). Given the absence of such a population, if stress relaxation in W films is the dominant source of phonon LEE events, the relaxation process must deposit a very small fraction of energy in the film electronic system, a possibility that we find improbable. In light of this work, other mechanisms, such as residual stress in the device holding scheme\cite{anthony-petersenStressInducedSource2022}, radiation induced defects in the device substrate\cite{DefectLEE}, or defects on the device surface, seem to more easily explain phonon (or ``shared'') LEE observations.

\section*{Acknowledgements}

This work was supported in part by DOE Grant DE-SC0019319, DE-SC0022354, and DOE Quantum Information Science Enabled Discovery (QuantISED) for High Energy Physics ( ). Work at Lawrence Berkeley National Laboratory was supported by the U.S. DOE, Office of High Energy Physics, under Contract No. DEAC02-05CH11231.

\section*{Refefences}
\bibliography{aipsamp}

\end{document}